\documentclass[]{spie}  %>>> use for US letter paper
\usepackage{amsmath,amsfonts,amssymb}
\usepackage{graphicx}
\usepackage[colorlinks=true, allcolors=blue]{hyperref}
\usepackage{svg}
\usepackage{subcaption}

\title{HEP digital micromirror devices for precision solar spectroscopy}

\author[a,b]{Christian Robles}
\author[a,b]{Suvrath Mahadevan}
\affil[a]{Department of Astronomy and Astrophysics, The Pennsylvania State University, 525 Davey Laboratory, 251 Pollock Road, University Park, PA, 16802, USA}
\affil[b]{Center for Exoplanets and Habitable Worlds, The Pennsylvania State University, 525 Davey Laboratory, 251 Pollock Road, University Park, PA, 16802, USA}

\authorinfo{Further author information: (Send correspondence to C.R.)\\C.R.: E-mail: crobles@psu.edu}

\begin{document} 
\maketitle

\begin{abstract}
We present the motivation and early tests for a novel solar instrument that will harness the new High Efficiency Pixel (HEP) Texas Instruments DLP801RE Digital Micromirror Device (DMD) as a reconfigurable spatial light modulator. 
This design enables real-time, dynamic configuration of the field of view for targeted spectroscopy of magnetically active regions and full-disk observations. Optical efficiency was validated through simulations and laser testing. Destructive window removal allowed for detailed structural analysis, confirming the elimination of central vias present in previous models. We measured a contrast ratio of 250:1, currently limited by the evaluation board’s duty cycle rather than the DMD itself. Furthermore, we successfully simulated artificial planetary transits, recovering depths ranging from gas giants to a 40 ppm rocky planet transit. These results demonstrate the HEP DMD’s potential for high-precision solar and exoplanetary science applications.

\end{abstract}

\keywords{Digital micromirror devices, astronomy, Extreme precision radial velocity}

\section{INTRODUCTION}

Digital Micromirror Devices (DMDs) are micro-electro-optical systems (MOEMS) originally developed by Texas Instruments for digital projection\cite{Hornbeck1987}. At their core, these devices consist of a grid of individual micromirrors controlled by underlying CMOS electronics. Through electrostatic repulsion, each mirror can be tilted into an ON or OFF state—typically between 10 to 15 degrees depending on the specific model—allowing for the precise spatial redirection of an optical signal. In the OFF state, the mirrors return to a mostly flat, unbounded position. Despite their small form factor (often less than one inch), DMDs are mechanically robust, a trait proven by decades of industrial stress testing\cite{douglass_2003} and their widespread adoption in commercial sectors ranging from cinema projectors to wavefront shaping\cite{hornbeck_1999,Popoff}.

In the context of astronomy, this programmable control allows DMDs to function as adaptive slits for multi-object spectrographs. Early implementations of this concept include IRMOS\cite{MacKenty} and RITMOS\cite{Meyer}. While DMDs were initially considered for the James Webb Space Telescope\cite{jwst2}, microshutter arrays\cite{moseley} were ultimately selected due to the higher contrast capabilities available at the time. However, modern commercial DMDs have since achieved significant improvements in both contrast and fill factor\cite{piot_2024}. These advancements enabled the development of SAMOS\cite{Smee2018,Smee2018b}, a multi-object spectrograph for the SOAR telescope, and have driven proposals for new large-pixel DMDs designed specifically for future space missions\cite{robberto_2025}.

For this work, we utilize the new High Efficiency Pixel (HEP) DMDs from Texas Instruments\cite{Dewa}, which offer distinct advantages for high-performance photometry. Specifically, we selected the DLP801RE model. This device features a 9-micron pitch and a large number of pixels (1920 $\times$ 1200), but crucially, it offers a maximum illumination rating of 40 W/cm$^2$—significantly higher than the standard DLP800RE. This high power handling is essential for solar observation, where thermal loads are substantial.
Furthermore, the HEP architecture improves optical throughput and contrast. The device incorporates a thinner Corning XG Eagle window with a visible light optimized anti-reflective (AR) coating, and the micromirrors possess an increased tilt angle of 14.5 degrees relative to earlier generations. While originally designed for LED projection, the high diffraction efficiency of these HEP devices makes them a promising candidate for use as a spatial light modulator (SLM). Validating this performance is a critical step toward advanced solar instruments to significantly advance the understanding of stellar variability—currently the most formidable obstacle to detecting Earth-twin exoplanets via the radial velocity method\cite{Crass,burt}.

\begin{figure*}[h!]
    \centering
    \includegraphics[width=\textwidth]{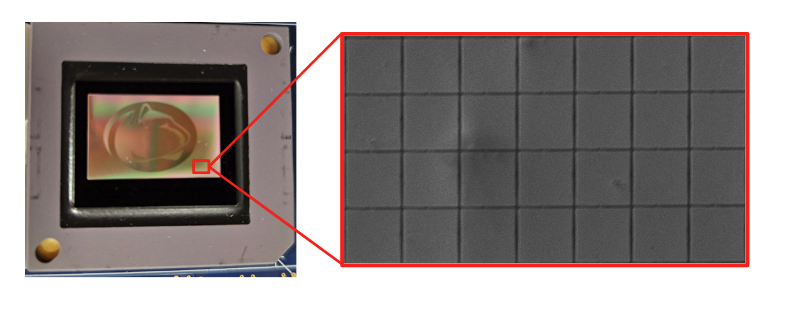}
    \vspace{2mm}
    \caption{Illustrative graphic showing the high efficiency pixel digital micromirror device from Texas Instruments. (left) Image of DMD with Penn State Logo. (right) SEM of HEP DMD surface showing the removal of the vias leading to a larger fill factor.}
\end{figure*}

\section{MOTIVATION}

The search for life beyond the Solar System has gained urgency following NASA’s selection of the Habitable Worlds Observatory (HWO) as its next flagship mission. Designed to directly image and characterize roughly 25 Earth-like exoplanets, HWO’s success depends on optimizing its mission lifetime for characterization rather than discovery\cite{hwo}. Consequently, suitable targets must be identified prior to launch. Given the low geometric probability of transits for long-period planets, the radial velocity (RV) method offers the most viable path for discovering these nearby candidates.
However, validating an Earth-twin around a G2V star requires detecting a reflex Doppler shift of merely $\sim$9 cm/s. 
The community has targeted this regime by developing Extremely Precise Radial Velocity (EPRV) instruments, including optical facilities like ESPRESSO, NEID, EXPRES, and MAROON-X, as well as near-infrared spectrographs such as HPF, SPIRou, NIRPS, and IRD\cite{esp,neid,expres,maroon,hpf,spirou,nirps,ird}.
Yet, despite instrumental precision reaching the $\sim$40 cm/s level\cite{Lin}, a fundamental bottleneck remains. Intrinsic stellar variability produces periodic and stochastic signals ranging from cm/s to m/s\cite{Cegla}, which can easily mask the planetary signature. Overcoming this stellar variability is currently the most formidable challenge facing the EPRV community\cite{Crass}. To overcome this, we require an instrument capable of using the resolved Sun as a laboratory to calibrate variability models. This paper explores two specific applications of such an instrument: mapping the RV impact of surface features and quantifying stellar contamination in transmission spectroscopy.

Solar RV signals vary across timescales ranging from the magnetic cycle (years) down to granulation (minutes). These fluctuations are driven by magneto-convective effects, where the interaction between the radiative core and the convective layer creates surface inhomogeneities such as spots, plage, and granulation\cite{gray2}. Granulation is particularly dominant; the contrast between hot upwelling plasma and cooler sinking lanes creates a net convective blueshift\cite{shp}.
When coupled with stellar rotation, these local surface velocities integrate to form complex global RV profiles. This is observed globally during the Rossiter-McLaughlin (RM) effect\cite{Queloz2000}, where a transiting body differentially blocks portions of the rotating stellar disk, revealing the local velocity structure. By utilizing a spatial light modulator to selectively mask regions of the solar disk, we can mimic this effect, directly probing the convective blueshift of specific solar features to build more robust models of stellar noise.

In addition to RV measurements, stellar surface features pose a challenge for transmission spectroscopy. Because transit depth is calculated as a ratio of in-transit to out-of-transit flux, any spatial inhomogeneity on the stellar surface introduces chromatic noise, known as the transit light source effect, which can mimic or obscure planetary atmospheric signals.
Simulations indicate that for an FGK star, this stellar contamination can introduce false signals of $\sim$6 ppt for Jupiter-sized planets and $\sim$20 ppb for 80 ppm Earth-analog transits\cite{rack19}. Consequently, NASA’s Exoplanet Exploration Program (SAG 21) has identified a critical need for resolved solar observations to benchmark these contamination models\cite{rack23}. An instrument capable of simulating a ''bare rock" transit across the resolved Sun would allow us to test approaches for disentangling stellar contamination from true planetary atmospheric signatures.

\section{EXPERIMENTAL SETUP}
To validate the High Efficiency Pixel (HEP) DMD for solar applications, we designed a benchtop experiment capable of evaluating photometric precision at the levels required for spatial control of the solar disk. We established a detection requirement of 25 ppm, which corresponds to the transit depth of a Mars-sized object. To recover these minute signals, the setup utilizes a highly stable broadband source, the HEP DMD as the optical modulator, and a high-precision integrating sphere detection system.

\subsection{Light Source Characterization} \label{light} 
\begin{figure*}[h!]
    \centering
    \includegraphics[width=.9\textwidth]{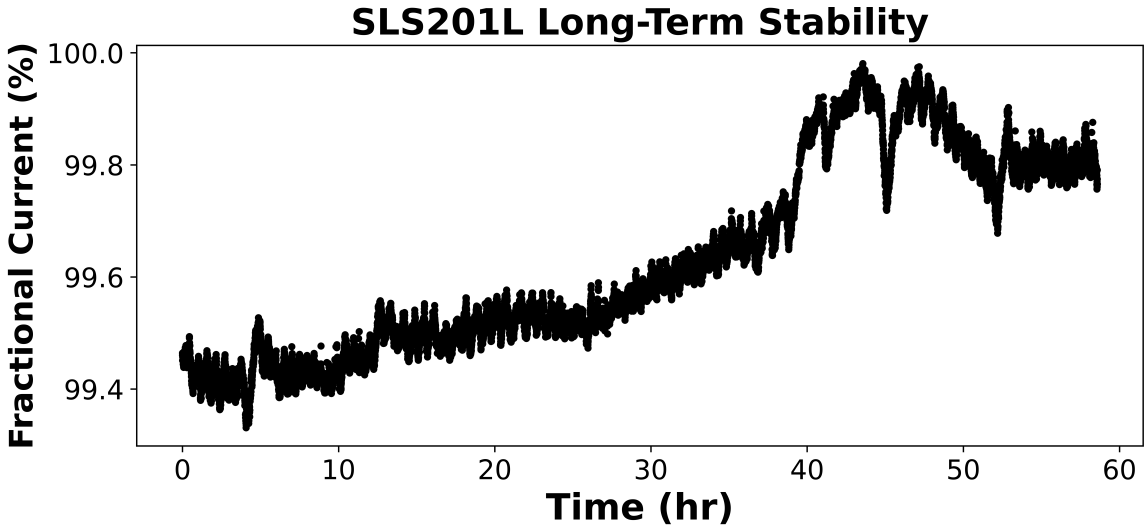}
    \vspace{2mm}
    \caption{Thorlabs stabilized QTH SLS201L shows high photometric stability over 60 hours.}\label{stable}
\end{figure*}

High-precision photometry requires a source with minimal intrinsic variability. We evaluated two quartz-tungsten-halogen (QTH) options: the AlphaBright and the Thorlabs (Model SLS201L) QTH lamps. Stability was measured over extended durations after a warm-up period.
The AlphaBright source exhibited a root-mean-square (RMS) fluctuation of 0.03\% (raw) and 0.014\% (after polynomial detrending) over 300 minutes. The Thorlabs source demonstrated similar stability with raw fluctuations of 0.07\% and a detrended RMS of 0.04\% over 400 minutes. Due to its superior thermal management for long-duration experiments, we selected the Thorlabs QTH. In a 60-hour stability test (see Fig.~\ref{stable}), this source achieved a final detrended RMS of 0.05\%.

\subsection{DMD Optical Configuration} The HEP DMD utilized the Texas Instruments (TI) DLP800RE evaluation module (Model DLP800REEVM) and the DLPC4430EVM controller board. To ensure the micromirror hinge axes were perpendicular to the optical table, the DMD was secured in a custom laser-cut acrylic mount oriented at a $45^{\circ}$ tilt. This assembly was affixed to a kinematic mount, providing independent pitch and yaw adjustments.
Alignment was established using a Helium-Neon (HeNe) laser defining the optical axis. The DMD pitch and yaw were adjusted until the primary reflection aligned with this axis, revealing the device's characteristic 2D diffraction pattern\cite{TI_laser,per_2022,Rice_2009}. The controller board allowed for pattern generation via a USB interface and manufacturer-supplied GUI.

\begin{figure*}[h!]
    \centering
    \includegraphics[width=\textwidth]{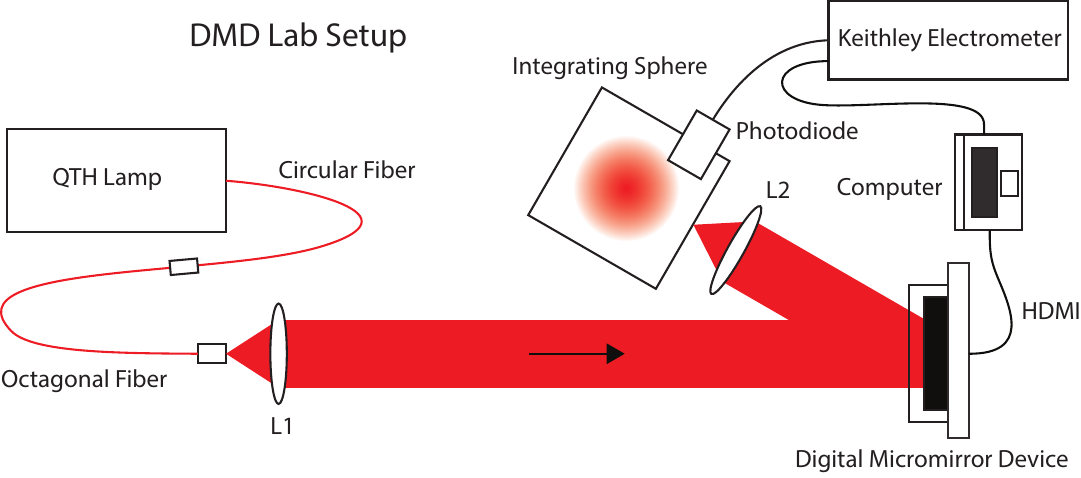}
    \vspace{2mm}
    \caption{Schematic of the DMD optical bench setup. The light from the Thorlabs QTH lamp is coupled into a circular to octagonal optical fiber which is then projected onto the DMD using Lens 1. The DMD directs a portion of light by  29 degrees to Lens 2 where the light is focused into a Thorlabs integrating sphere.}
\end{figure*}

\subsection{Photometric Measurement System} \label{measure} The DMD directs light into two discrete states ($\pm14.5^{\circ}$). We utilized the ON channel for photometric measurements and the OFF channel for alignment verification, projecting a magnified image of the DMD surface to inspect the device's ''pond-of-mirrors" border region.
To mitigate photometric errors caused by beam spatial inhomogeneity, light from the ON state was coupled into a Thorlabs integrating sphere (Model 2P3) using a Hastings triplet lens (Thorlabs TRH254-040-A-ML) at $f/2.25$ focal ratio. The integrated flux was detected by a silicon photodiode (Thorlabs SM05PD1A) housed in a recessed port to ensure multiple reflections before detection.
The photodiode operated in photovoltaic mode to maximize linearity and minimize dark current, which was measured at a median of 0.03 pA (compared to signal levels of $\sim$10s of nA). The resulting photocurrent was digitized by a Keithley 6514 electrometer with 6.5-digit precision. Data acquisition was automated using custom Python scripts interfacing with the electrometer via a GPIB-to-USB connection and National Instruments drivers.

\begin{figure*}[h]
    \centering
    \begin{subfigure}[t]{0.45\textwidth}
        \centering
        \includegraphics[width=\linewidth]{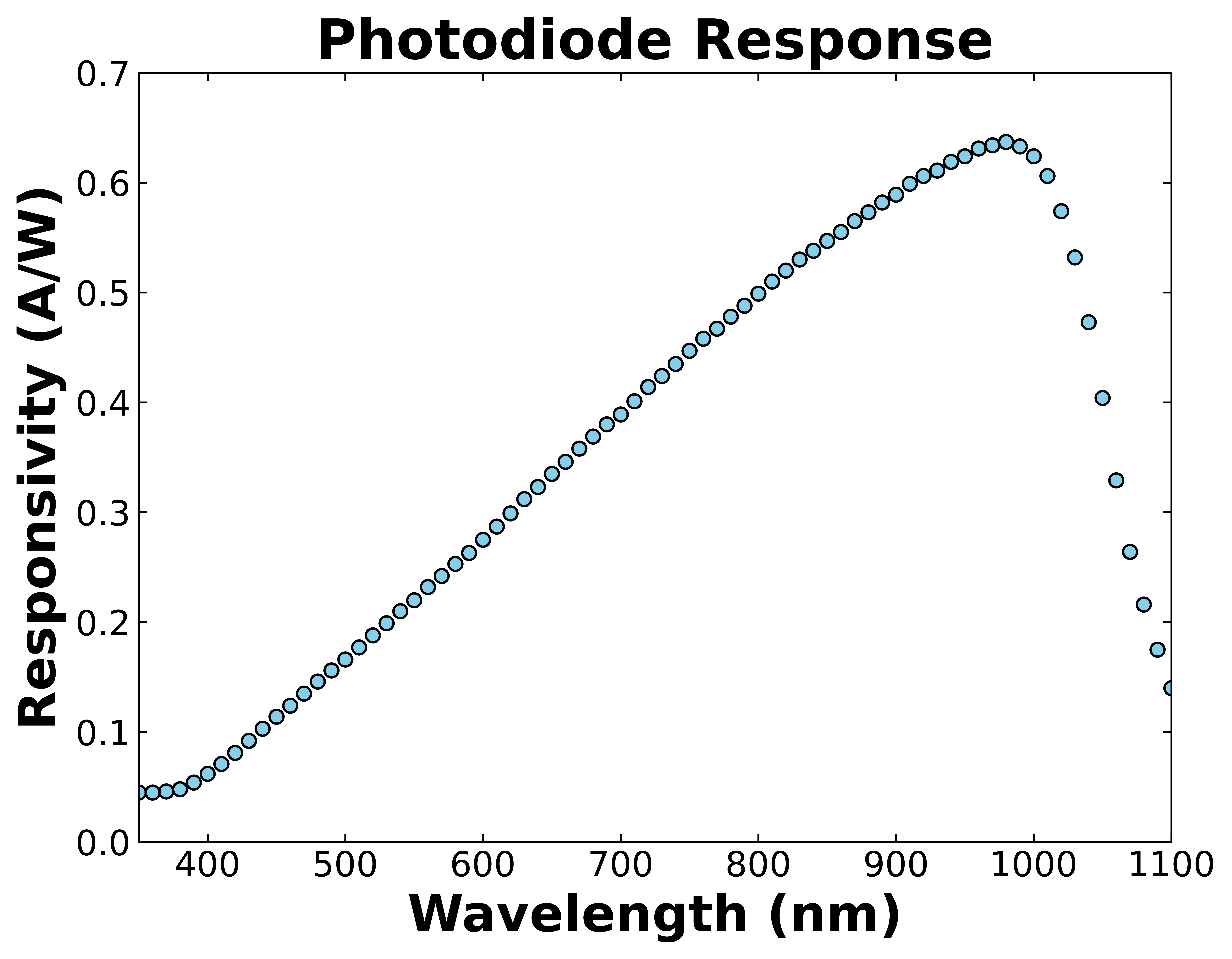} 
        \caption{FDS100 photodiode responsivity in the optical from Thorlabs\cite{Thorlabs_FDS100_Specs}.} \label{fig:timing1}
    \end{subfigure}
    \hfill
    \begin{subfigure}[t]{0.45\textwidth}
        \centering
        \includegraphics[width=\linewidth]{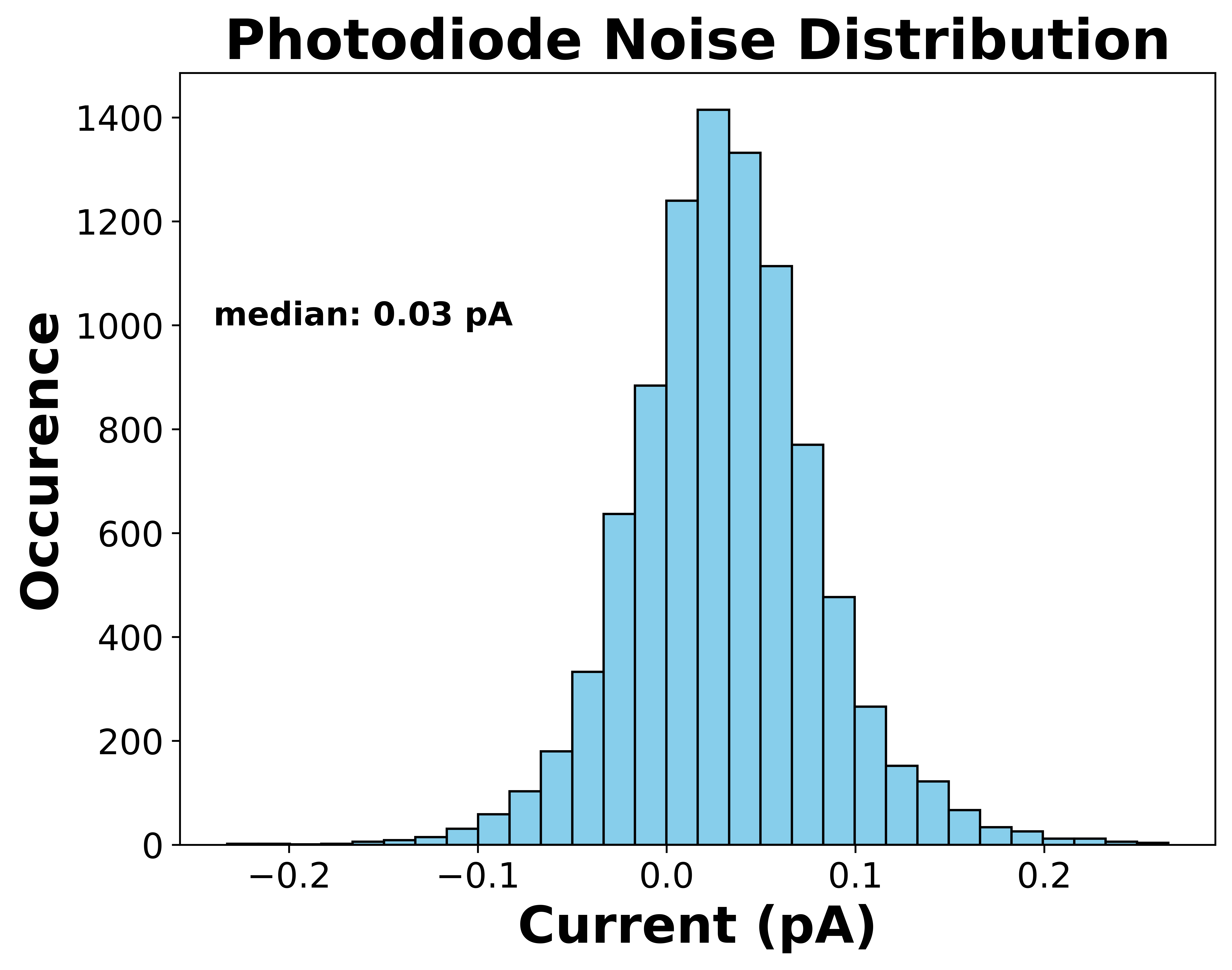} 
        \caption{Measured photodiode noise.} \label{fig:timing2}
    \end{subfigure}
    \caption{}
\end{figure*}

\section{CHARACTERIZATION OF THE HEP DMD} As the High Efficiency Pixel (HEP) architecture is a recent introduction\cite{Dewa}, independent verification of its optical performance is necessary for high-precision applications. We characterize these devices through a combination of analytical modeling, manufacturer specifications, and empirical testing.

\subsection{Optical Efficiency} The total optical efficiency of the device is defined by the window transmission, micromirror fill factor, diffraction efficiency, and surface reflectivity. Because incident light must pass through the protective window twice (once before and once after reflection), the efficiency is modeled as: 
\begin{equation}
    E_{\text{DMD}} = T^{2}_{\text{window}} \times F \times E_{\text{diffraction}} \times R_{\text{mirror}}
    \label{op_eff}
\end{equation}

where $T_{\text{window}}$ is the single-pass transmission, $F$ is the fill factor, $E_{\text{diffraction}}$ is the diffraction efficiency, and $R_{\text{mirror}}$ is the aluminum reflectivity. Notably, the HEP architecture increases the fill factor to 97\% (up from 92\% in previous generations) by filling the vias to maximize the reflective surface area.

\begin{figure*}[h!]
    \centering
    \includegraphics[width=.5\textwidth]{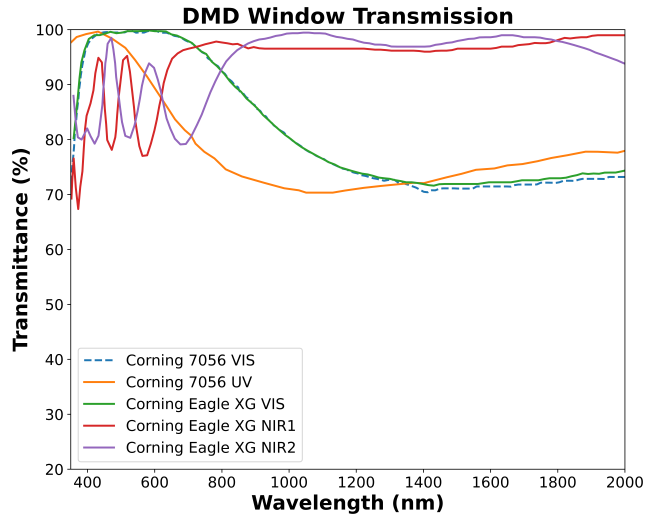}
    \vspace{5mm}
    \caption{Window transmission curves for common windows on Texas Instruments DMDs\cite{TI_window}. The Corning Eagle XG VIS is used on the HEP DMDs.}\label{win}
\end{figure*}

\subsubsection{Window Transmission and Modification} 
\begin{figure*}[h!]
    \centering
    \includegraphics[width=.5\textwidth]{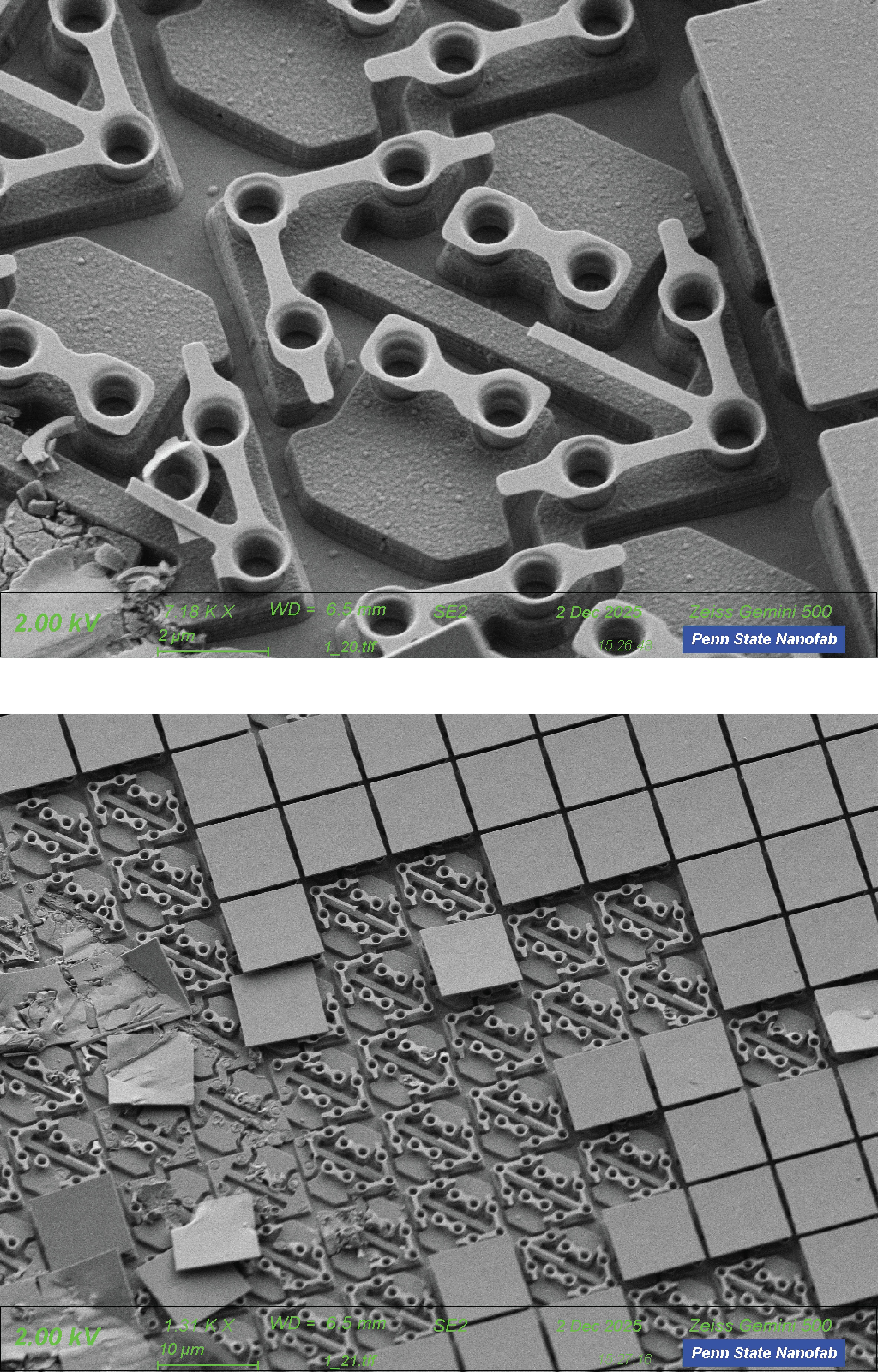}
    \vspace{5mm}
    \caption{HEP DMD window was destructively removed to study the wafer level structure. SEMs of the surface of the DMD. The substructure of each DMD pixel is common to previous DMD models. The surface of the micromirrors show the filled vias.}\label{destroy}
\end{figure*}

The DMD is hermetically sealed with Corning Eagle XG glass featuring a double-sided anti-reflective (AR) coating optimized for visible wavelengths (400–700 nm). While these coatings are highly efficient at normal incidence (Fig.~\ref{win}), they limit the device's utility in the NIR. To evaluate the feasibility of customizing the spectral range, we investigated window removal techniques.
Previous generations allowed for successful window replacement\cite{gallego_2023} although it is unclear if this same approach would be effective on the HEP packaging. Our attempts to remove the HEP window via thermal stress using a hot-air rework station were unsuccessful due to the robustness of the encapsulant. A subsequent destructive removal via mechanical puncture allowed for an internal inspection but resulted in significant surface contamination and mirror damage. We obtained a Scanning Electron Microscopy (SEM) of the resulting structure (Fig.~\ref{destroy}) that reveals the underlying hinge architecture, confirming that window removal on this package remains a high-risk procedure.

\subsubsection{Diffraction Efficiency} 
\begin{figure*}[h!]
    \centering
    \includegraphics[width=.6\textwidth]{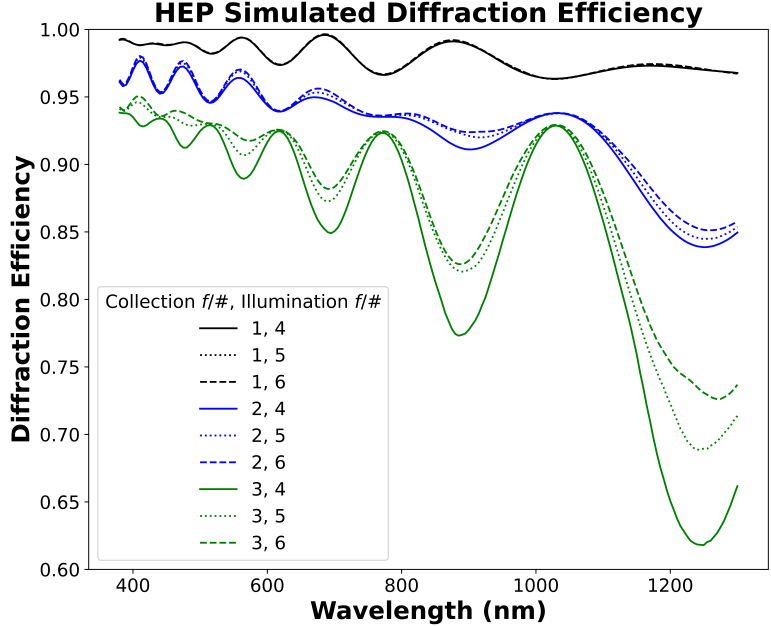}
    \vspace{5mm}
    \caption{Simulations of the effect of illumination and projection solid angle on the diffraction efficiency. This was run in the Texas Instruments DMD Diffraction Efficiency Calculator with values for the HEP DMDs. The figure is colored by collection F-number to show the more angular area is collected leads to a higher diffraction efficiency.}\label{sim_fnum}
\end{figure*}

\begin{figure*}[h!]
    \centering
    \includegraphics[width=.8\textwidth]{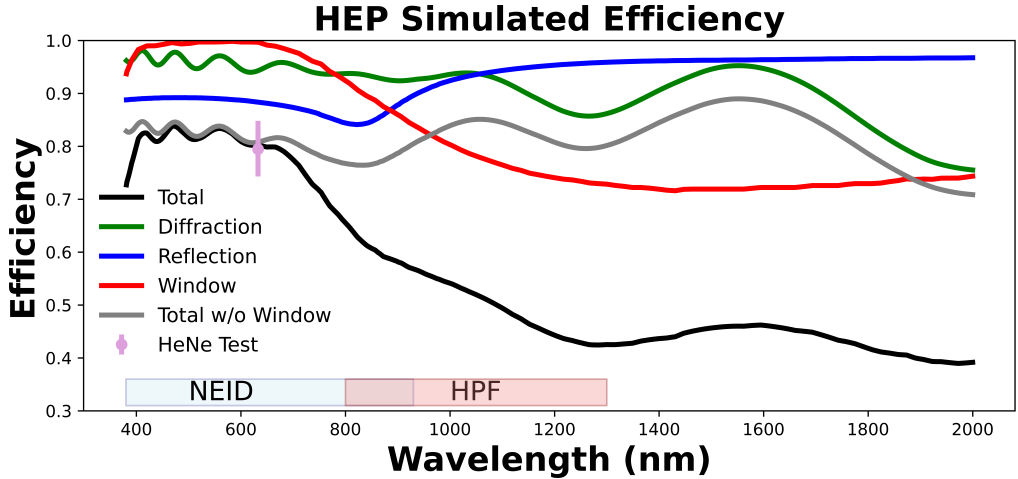}
    \vspace{5mm}
    \caption{The overall calculated efficiency for the HEP is strongest in the optical and suffers in the NIR from the double transmission through the window with a visible AR coating. A line is shown for this total without the window to show the theoretical limit if the window is replaced for the HEP DMDs use in the NIR. The measured HeNe test point is in good agreement with the simulated value although more wavelength coverage would aid in validating this calculation. On the bottom shows the wavelength ranges for the NEID and HPF spectrometers to display wavelengths of interest to the EPRV community.}\label{total_eff}
\end{figure*}

Because the DMD acts as a 2D diffraction grating, its efficiency is chromatic and highly dependent on the system’s F-number. We simulated the impact of illumination and collection solid angles using provided TI DMD Diffraction Efficiency Calculator\cite{TI_diff_calc}. Our simulations (Fig.~\ref{sim_fnum}) indicate that larger collection solid angles (lower F-numbers) capture more diffraction orders and increase efficiency.
To validate these simulations, we measured a single-point diffraction efficiency using a HeNe laser. The empirical results (Fig.~\ref{total_eff}) showed strong agreement with the model. Future work will expand this to broadband measurements to characterize the full spectral diffraction envelope.

\subsection{Contrast Measurement}
Contrast—defined as the ratio of light captured in the ON channel to the scattered light measured when mirrors are in the OFF state—was evaluated using a time-series modulation. The DMD was commanded through a binary sequence, alternating between all-active (ON) and all-inactive (OFF) states under broadband illumination from the Thorlabs QTH lamp. To isolate the device performance from source instability, the resulting light curve was phase-folded and the ON state baseline was fit with a polynomial. Normalizing the time-series by this fit provided a corrected fractional current. Under active control, we measured a contrast ratio of 250:1. However, when the DMD was completely unpowered—forcing the mirrors into an unbounded flat state—the contrast improved to 400:1. While this 400:1 ratio is closer to the 1600:1 reported in literature\cite{Dewa}, the significant discrepancy between the unpowered and active OFF states suggests that the manufacturer-supplied controller board is not holding the micromirrors statically.

\begin{figure*}[h!]
    \centering
    \includegraphics[width=.6\textwidth]{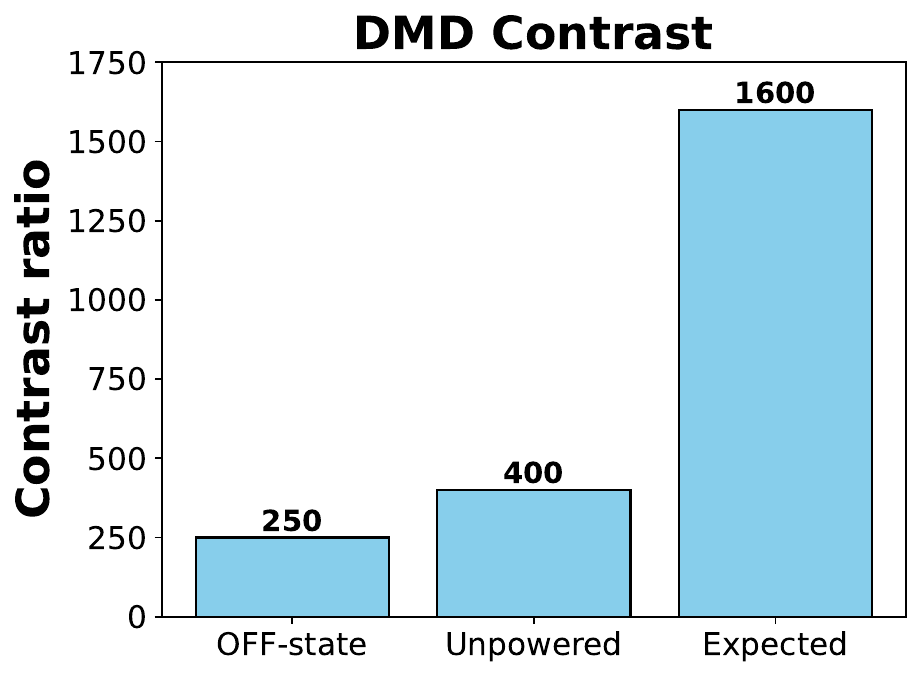}
    \vspace{5mm}
    \caption{The contrast measurements compared to the expected value from Ref.~\citenum{Dewa}. The OFF-state is the normal contrast measurement while the DMD was also tested when it was unpowered. This unpowered measurement should be lower than the state in which the micromirrors are directing the light in the opposite direction. This not the case in our setup as we had issues of eliminating the duty cycle imposed by the manufacturer supplier electronics board and software.}
\end{figure*}

\section{TRANSIT TESTING}
\begin{figure*}[h!]
    \centering
    \includegraphics[width=.8\textwidth]{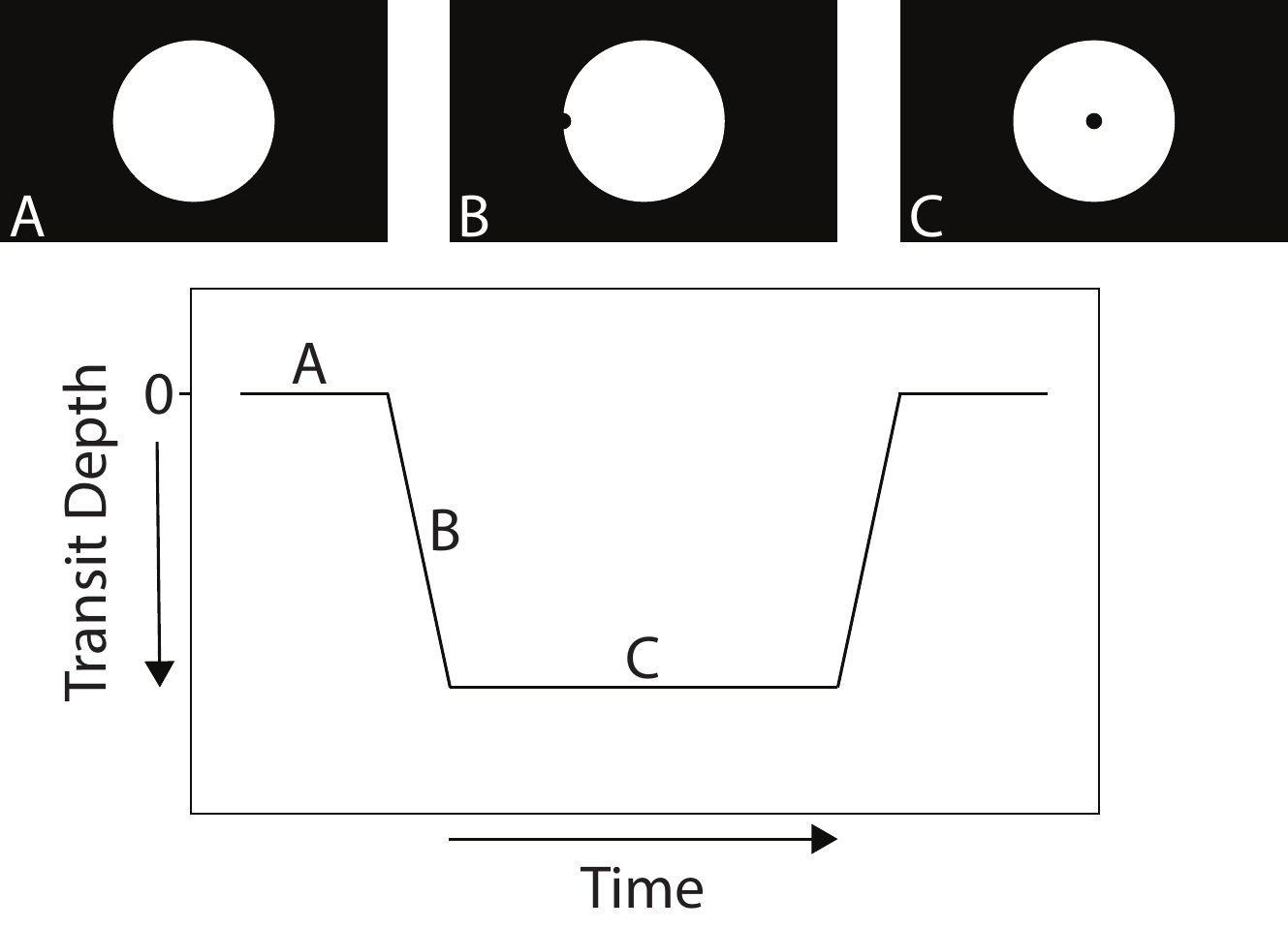}
    \vspace{2mm}
    \caption{(top) Images used on the surface of the DMD to simulate gas giant transits. `A' is out-of-transit, `B' is ingress, ie. entering the stellar disk, and `C' is in-transit.}\label{tran}
\end{figure*}

A planetary transit occurs when an exoplanet occults a portion of its host star’s disk, resulting in a wavelength-dependent decrease in the observed integrated flux as illustrated in Fig.~\ref{tran}.

\subsection{Illumination and Spatial Uniformity} To simulate high-precision transits, the illumination across the DMD must be spatially uniform. Initial tests using standard circular fibers revealed peak-to-peak variations across the central active portion of the DMD. To mitigate this, we employed a spatial scrambling technique using a 400 $\mu$m circular multimode patch fiber coupled into an octagonal fiber. Octagonal fibers facilitate near-perfect near-field scrambling by suppressing the spiral modes common in circular fibers\cite{sam_arp}. The near-field of the octagonal fiber was imaged onto the DMD surface, providing a highly uniform ``stellar" disk for modulation.

\subsection{Synthetic Planetary Transit Generation}
To demonstrate the precision control of the DMD and its potential application in astronomy, we physically simulated planetary transits on the DMD and photometrically recovered the resulting light curves using the experimental setup.
The DMD was controlled via an HDMI connection, mirroring the display of a laptop at the DMD's native resolution ($1920 \times 1200$ pixels). This configuration allows the color of each screen pixel to individually control a corresponding micromirror: full white maps to the ON configuration, and black maps to the OFF state. Only full white or black were used. 

The DMD was illuminated with a near field image of the octagonal fiber from the Thorlabs QTH Lamp.
A GIF was generated using the Python Pillow package to create the base stellar flux, represented by a large central white circle. The star was set to a radius of $400$ pixels, covering $66\%$ of the DMD along its short axis. 
Planetary transits were simulated by introducing a black circle to each frame of the GIF, blocking a fraction of the total light. The planet would enter the star and proceed across the diameter. The GIF generation code included configurable parameters for the transit period, the ratio of in-transit to out-of-transit time, the planet's radius (in pixels), and the frames per second (FPS). The planetary radius directly determines the desired transit depth when the size of the star is constant. For a typical simulation, we tested a Jupiter-sized transit with a $40$ pixel radius, a $50$-second period, a $50\%$ transit duration ratio, and $30$ FPS. 

\subsection{Light Curve Reduction and Stabilization}
After generating the time series, the data underwent several reduction steps:
\begin{enumerate}
    \item Phase Folding: The light curve was phase-folded using the known period of the synthetic transit.
    \item Drift Correction: An out-of-transit mask was applied to isolate baseline data points. A spline was then fit to these out-of-transit data points across the full time series to model and correct for the drift present in the light source.
    \item Fractional Flux Calculation: The corrected fractional photocurrent was calculated by dividing the measured photocurrent by the baseline-fitted spline.
    \item Binning: The final phase-folded fractional light curves were binned every $100$ data points to reduce noise.
\end{enumerate}

\subsection{Transit Detection and Summary}
\begin{figure}[htbp]
     \centering
     \begin{subfigure}[b]{0.45\textwidth}
         \centering
         \includegraphics[width=\textwidth]{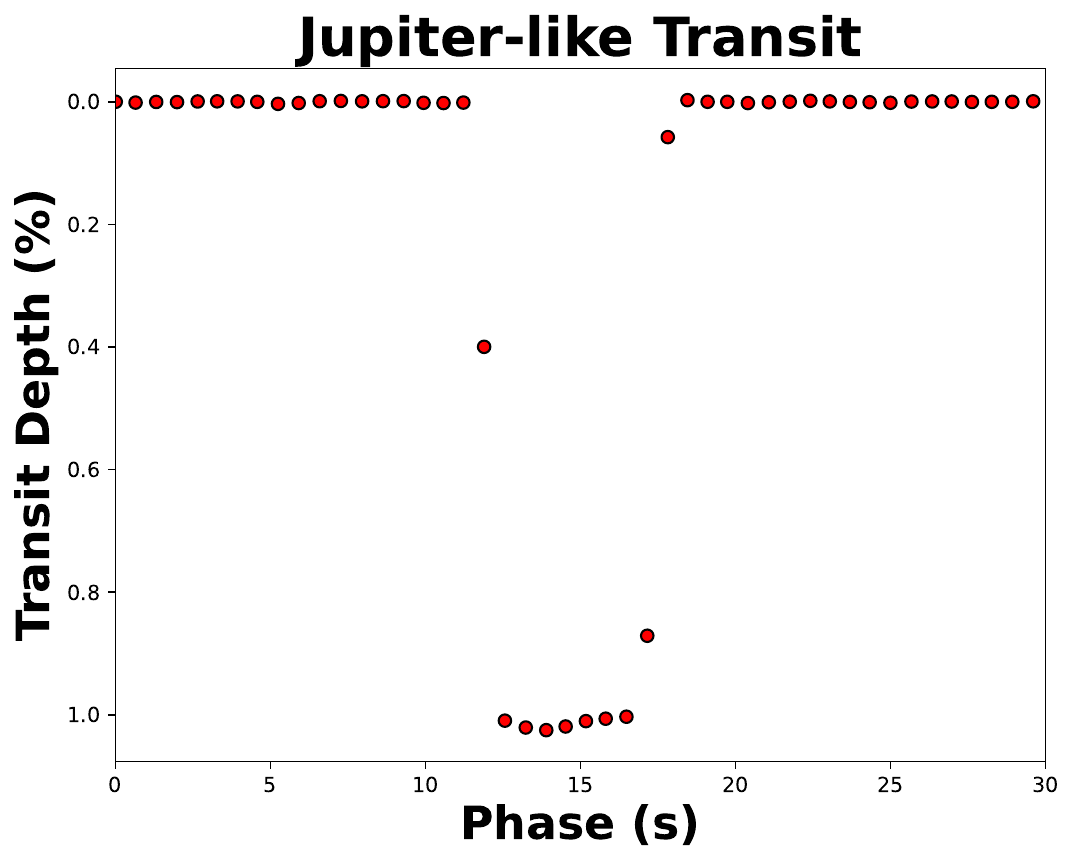}
         \label{fig:plot1}
     \end{subfigure}
     \hfill
     \begin{subfigure}[b]{0.45\textwidth}
         \centering
         \includegraphics[width=\textwidth]{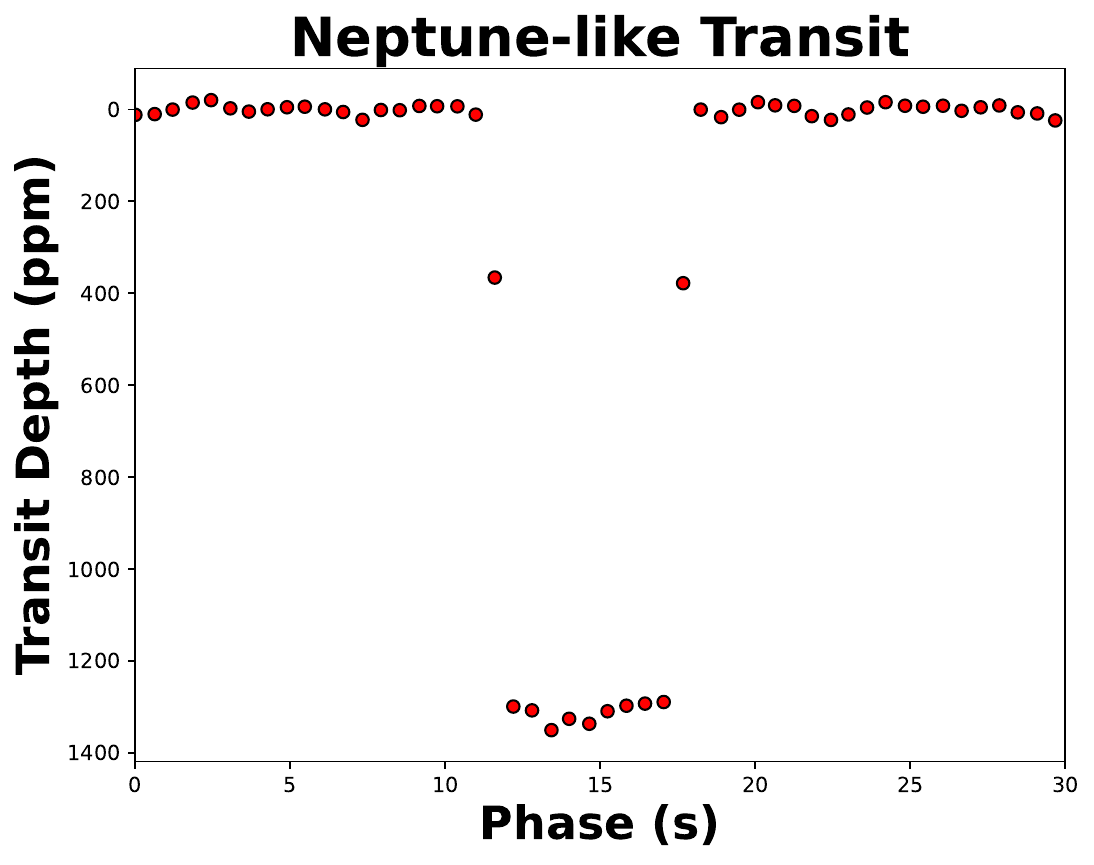}
         \label{fig:plot2}
     \end{subfigure}

     \vspace{0.5cm}

     % --- Second Row ---
     \begin{subfigure}[b]{0.45\textwidth}
         \centering
         \includegraphics[width=\textwidth]{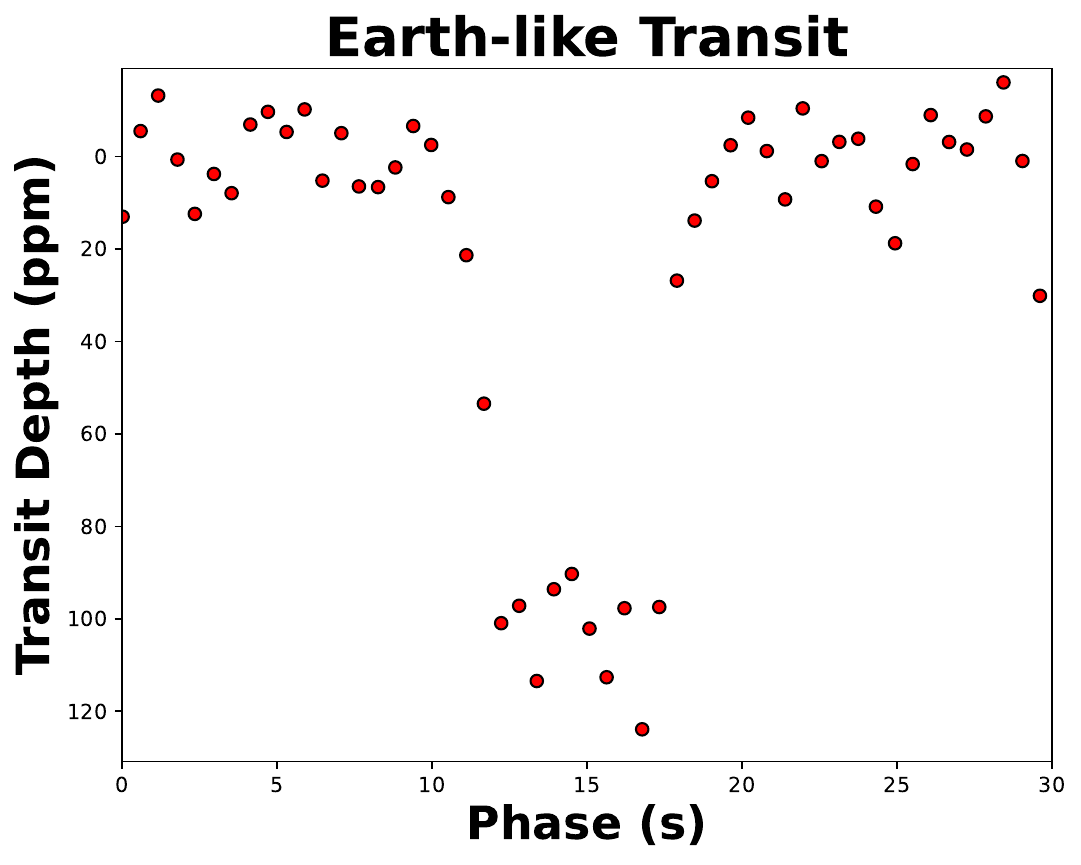}
         \label{fig:plot3}
     \end{subfigure}
     \hfill
     \begin{subfigure}[b]{0.45\textwidth}
         \centering
         \includegraphics[width=\textwidth]{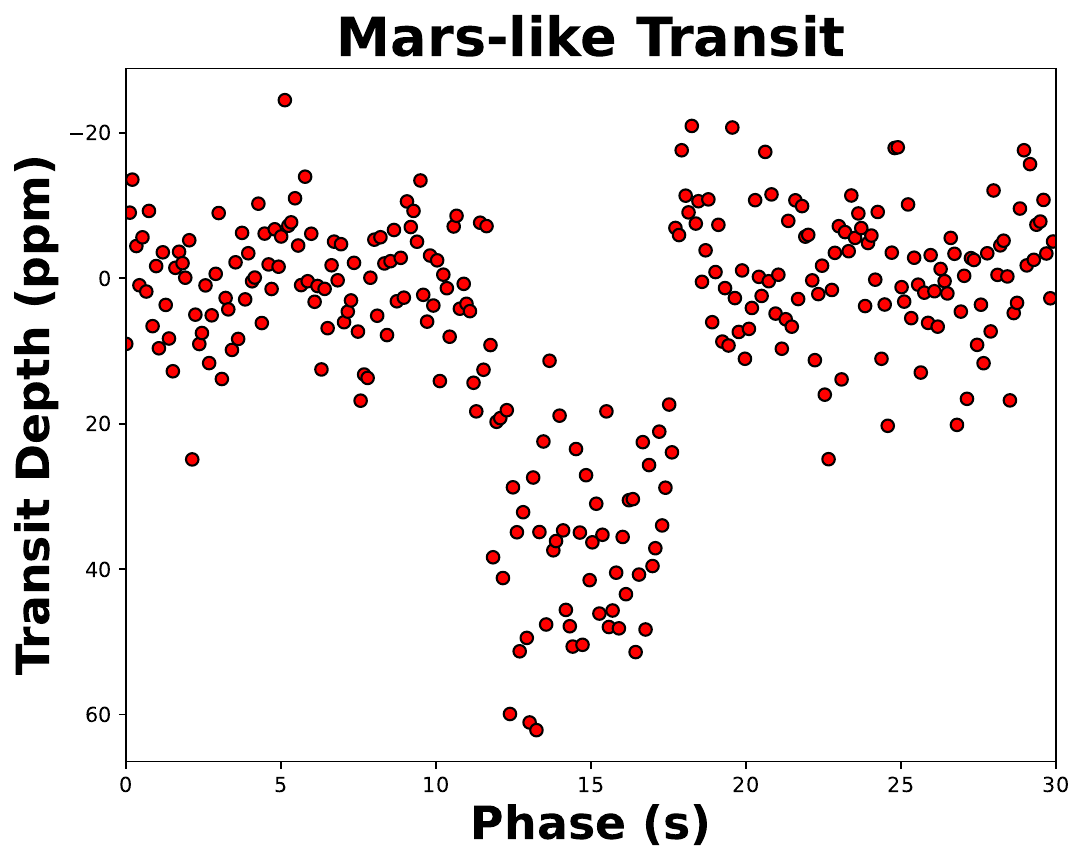}
         \label{fig:plot4}
     \end{subfigure}
     
     \caption{Phase-folded and binned (N=100) photometric data for four synthetic planetary sizes. The transits were produced by modulating the DMD surface to mask a uniform stellar disk. Detrending was applied to isolate the signal from intrinsic lamp variability.}
     \label{big_plot}
\end{figure}

We successfully simulated and recovered synthetic transit signals for a range of planetary scales, from gas giants to terrestrial analogs. As shown in Fig.~\ref{big_plot}, the signals for Jupiter, Neptune, and Earth were recovered from approximately 25 phase-folded transits. To achieve the higher signal-to-noise ratio (SNR) required for the Mars-analog, we folded approximately 150 transits.

A key constraint in these simulations is the discretized nature of the DMD grid. Because the stellar and planetary disks are composed of individual micromirrors, the achievable transit depth is limited by the spatial quantization of the pixels. While our target depth for the Mars-sized occultor was 25~ppm, the quantization of the Mars radius produced a transit depth of 40~ppm. Despite this quantization limit, the 40~ppm signal was accurately recovered. The recovery of these transits validate the precision of control required to generate artificial transits on the sun to test atmospheric transmission retrieval methods with no planetary atmospheres.

\section{CONCLUSION AND FUTURE WORK} This work demonstrates that the new High Efficiency Pixel (HEP) DMD architecture can achieve the photometric precision required for the next generation of astronomical instrumentation. Through benchtop experiment, we successfully recovered synthetic planetary transits down to the Mars-like transit ($\sim$10s ppm), validating the device's capability for the high-stability spatial modulation required in solar and exoplanetary science.

However, our characterization also identified critical constraints. The measured contrast discrepancy between unpowered (400:1) and active (250:1) states suggests that standard commercial evaluation boards may limit the device's theoretical performance. For applications requiring extreme contrast, the development of dedicated, high-stability control electronics will be a necessary next step.

Despite these hardware-specific challenges, the HEP DMD remains a highly promising candidate for a spatial light modulator in multi-object spectrographs and solar observing instruments. Future work will focus on integrating this technology into a prototype solar instrument designed to mask specific solar regions for simultaneous spectroscopy and imaging. By providing configurable access to solar regions, these DMD-based instruments could play a crucial role in pushing radial velocity detection limits toward the terrestrial exoplanet regime.

\acknowledgments
This material is based upon work supported by the National Science Foundation Graduate Research Fellowship Program under Grant No. DGE1255832. Any opinions, findings, and conclusions or recommendations expressed in this material are those of the author(s) and do not necessarily reflect the views of the National Science Foundation. This work was partially supported by funding from the Center for Exoplanets and Habitable Worlds. The Center for Exoplanets and Habitable Worlds is supported by the Pennsylvania State University and the Eberly College of Science. Gemini AI (Google) was used for technical editing and structural organization assistance.    

We acknowledge help from Ishan Rana, who, took the SEM images of the surface of the DMD. Larry Ramsey assisted in the concept planning for experiments and future use in an instrument. Kristo Ment assisted with preliminary transit recovery and provided technical expertise for analyzing transit data. Steve Smee shared his experience and insight into DMD uses in astronomy and in help in trying to deconstruct the package for window replacement.

% References
\bibliography{ref} % bibliography data in report.bib
\bibliographystyle{spiebib} % makes bibtex use spiebib.bst

\end{document}